\documentclass[10pt,twocolumn]{article}
\usepackage[letterpaper,margin=0.75in,top=0.8in,bottom=0.9in]{geometry}
\usepackage[T1]{fontenc}
\usepackage[utf8]{inputenc}
\usepackage{times}
\usepackage{microtype}
\usepackage{graphicx}
\usepackage{booktabs}
\usepackage{amsmath}
\usepackage{url}
\usepackage{xcolor}
\usepackage[colorinlistoftodos,textsize=footnotesize]{todonotes}

\usepackage[numbers,sort&compress]{natbib}
\usepackage[colorlinks=true,linkcolor=blue!60!black,citecolor=blue!60!black,urlcolor=blue!60!black]{hyperref}
\usepackage{caption}
\title{\LARGE Before the Ticket: Grounding HPC User Support in Documentation and Resolved Tickets with an Agentic LLM Assistant}
\author{%
Elham J. Barezi,
Primus C Kabuo,
Niveditha Nerella\\
\texttt{<ejebalba,ckabuo@purdue.edu>}\\
Rosen Center for Advanced Computing, Purdue University, West Lafayette, IN, USA
}

\begin{document}
\maketitle

\begin{abstract}
Most questions that reach an HPC help desk have already been answered, in the center's user guides or in a ticket someone else filed last semester; users still file new tickets because the guides are hard to navigate and the ticket history is not theirs to search. We present an assistant that answers from both sources: an orchestrator calls two tools over the Model Context Protocol, retrieval-augmented generation (RAG) over the live documentation and a Llama-3-8B model fine-tuned with QLoRA on 11{,}244 query--resolution pairs mined from historical tickets, and merges their answers. Every model runs on-premises or on the university's hosted LLM service, because ticket data may not leave the institution. On 400 held-out tickets from Purdue's Rosen Center for Advanced Computing, ticket-disjoint and free of near-duplicates from training, and with two judges from model families that generated none of the answers, fine-tuning on resolutions rewritten into self-service guidance is the clear winner on agreement with RCAC's resolutions (BERTScore F1 0.06$\rightarrow$0.27 over the base model) and is judged most correct on average. The end-to-end system does not improve on its tools, although they are complementary: choosing the better answer per question would exceed either (judged correctness 3.41 vs.\ 3.20, an oracle bound), while the merge follows the documentation tool in 78\% of answers. Returning one tool's answer instead recovers the fine-tuned tool's quality (3.19 vs.\ 2.66), but neither a learned selector nor an LLM verifier beats that tool alone.
\end{abstract}

\section{Introduction}
\label{sec:intro}

A research computing center is a strange kind of product: thousands of nodes, dozens of file systems and schedulers, hundreds of software modules, and a user base of physicists, biologists, and economists who would rather not learn any of it. Most of them never took a course on it; HPC has only recently entered undergraduate curricula \citep{raj2020}. When a job fails or a module will not load, they file a ticket. One large public-university center logged more than 20{,}000 tickets and nearly 95{,}000 messages between 2017 and 2023, with an average resolution time of 11.8 business days \citep{rodenbeck2023}; help-desk staff across industries report spending 60--70\% of their time on problems they have solved before \citep{gonzalez2005}.

The frustrating part is that the answers usually exist. Every center maintains user guides. Every ticketing system is, in effect, a growing FAQ written by the support team one reply at a time. But the two live in different places and hold different kinds of knowledge. Documentation is authoritative and kept current, but it describes the system as designed; it is silent on the specific error a user is staring at and on the shortcuts staff reach for in practice. Ticket history captures exactly that tacit, experiential knowledge, but it is not accessible to users (tickets contain other people's data), it ages as the system changes, and nobody has time to curate it. A good support answer often needs both: what the manual says and what actually fixed it last time.

Large language models make it possible to put both sources behind one conversation. The question we address is how to combine them so that answers stay grounded, under the practical constraint that ticket data cannot leave the institution and so the models must be small enough to run on local or institutionally hosted hardware. Prior LLM work on IT support has mostly automated the pipeline \emph{after} a ticket is filed, classification, routing, escalation, FAQ generation \citep{zangari2023, paket2024, liu2025, joslin2025}. We instead intervene before the ticket.

\paragraph{Contributions.}
\begin{enumerate}\itemsep2pt
  \item An \textbf{agentic architecture} in which an orchestrator routes queries to two MCP-exposed tools, RAG over official documentation and a fine-tuned model over resolved tickets, and synthesizes their outputs with an explicit precedence rule and source attribution (Section~\ref{sec:system}).
  \item A \textbf{pipeline that turns raw ticket threads into assistant-safe training data}: LLM extraction of query--resolution pairs, rewriting of staff-performed resolutions into self-service guidance, filtering of answers that claim actions, residual PII masking, and a ticket-level, near-duplicate-free split (Section~\ref{sec:data}).
  \item A \textbf{controlled evaluation of five systems on the same 400 held-out tickets} (base model, fine-tuned model, RAG over documentation, RAG over tickets, end-to-end system) with reference metrics, two independent LLM judges, and paired bootstrap intervals (Section~\ref{sec:eval}). Fine-tuning wins on agreement with staff resolutions; the two tools are complementary, but the synthesis does \emph{not} improve on them, and selectors developed on validation data recover the fine-tuned tool's quality without exceeding it; judge choice and answer length change the picture enough that neither can be left out of a report.
\end{enumerate}

\section{Setting and Data}
\label{sec:data}

\subsection{The center}
The work was carried out at Purdue's Rosen Center for Advanced Computing (RCAC), which runs several shared clusters (Anvil, Bell, Negishi, Gilbreth, Gautschi, Geddes, Scholar) for researchers, students, and research software engineers. Support flows through a central service desk; staff consult public user guides, internal pages, and the ticket history, which RCAC already indexes with NLP in its TicketHub platform \citep{rodenbeck2023}. The user guides are public. The tickets are not.

\subsection{Knowledge sources}
\textbf{Documentation}: every page on RCAC's documentation portal plus a legacy knowledge base that still hosts cluster- and storage-specific guides, 849 pages in the evaluated crawl. \textbf{Tickets}: of the three systems in TicketHub (Table~\ref{tab:sources}) we use the two active ones, TeamDynamix and Anvil Jira, giving 15{,}268 tickets from 2023 onward. The larger legacy Footprints corpus was excluded because much of it concerns retired hardware and documentation that no longer exists.

\begin{table}[t]
\centering\small
\caption{Historical ticket sources.}
\label{tab:sources}
\begin{tabular}{lccc}
\toprule
System & Period & Tickets & Used \\
\midrule
Purdue Footprints (legacy) & 2016--2023 & 19{,}714 & No \\
Purdue TeamDynamix & 2023--present & 9{,}630 & Yes \\
Anvil Jira & 2023--present & 5{,}732 & Yes \\
\bottomrule
\end{tabular}
\end{table}

\subsection{Turning threads into training data}
\label{sec:pipeline}
A ticket's \texttt{customernote} field is the whole conversation, problem statement, clarifying questions, staff replies, signatures, quoted e-mail, Jira markup. Five steps turn it into something a model can learn from and be tested on.

\textit{Clean and pseudonymise.} Records with no note were dropped. Submitter and assignee fields were mapped to stable pseudonyms so repeat-user and workload analyses remain possible without identities. A fixed pipeline then stripped hyphen runs, HTML, issue-tracker markup, and non-ASCII residue; normalized Unicode and whitespace; and replaced e-mail addresses, and names matching the submitter and assignee lists, with \texttt{[EMAIL]} and \texttt{[NAME]} placeholders, which keep sentences grammatical.

\textit{Extract query--resolution pairs with an LLM.} Threads are too varied for regular expressions, so an LLM produced a JSON object with the user's \emph{query} in the first person and the \emph{resolution} as a direct instruction, keeping commands, paths, module names, error strings, and URLs verbatim, dropping greetings and metadata, and returning an empty resolution when the ticket was unresolved. We compared three extractors on 50 tickets, Gemma~3 (27B) with a prompt, Llama-4, and the Mistral-7B model that \citet{joslin2025} fine-tuned for this exact task, using three independent LLM judges (ChatGPT (GPT-4o mini), Gemini (3.5 Flash) and Claude Opus 4.8, through their chat interfaces). Prompted Gemma~3 matched the purpose-built Mistral model (issue extraction 9.0 vs.\ 8.0; equal hallucination-risk and RAG-suitability scores) and clearly beat Llama-4, so we used it without any fine-tuning. Extraction ran on Purdue's on-premises GenAI Studio so proprietary data stayed inside university infrastructure; it was parallelised with retries and checkpoints, and tickets that came back empty were retried up to twice more. Removing those the model judged unresolved (1{,}153) and those that still failed (783) left 13{,}332 pairs.

\textit{Rewrite staff actions; remove non-answers.} This step turned out to matter. Many resolutions say what the \emph{engineer} did: ``I've added you to the \texttt{standby} queue and bumped your scratch quota to 200~GB.'' Others record a passing state: ``Geddes is under maintenance; we'll update you.'' An assistant trained on these will claim to have granted access it cannot grant and assert outages it cannot see. Every resolution was passed through a rewriting prompt for Gemma~3, which left self-service answers essentially unchanged and turned staff actions into user-facing guidance that keeps the technical content and explains how to \emph{request} the privileged change (8 rewrites failed, leaving 13{,}324 pairs). Rewriting does not catch everything, so answers that still claim an action or report a transient state were removed with regular-expression rules. A first set of rules (``has been added'', ``you should now be able to'', ``are investigating'') missed about 3\% of such answers, for example ``The data depot should now be accessible'' and ``Your purchased 2TB quota has been applied''; the extended rules removed 632 pairs in total (4.7\%), leaving 12{,}692. The rules are deliberately strict and also catch some useful answers.

\textit{Mask residual PII.} The upstream pseudonymisation missed identifiers that a fine-tuned model can learn and repeat. A second pass, with the same placeholders, masked personal e-mail addresses (published support addresses such as \texttt{rcac-help@purdue.edu} are kept), phone numbers, people's names (from the submitter and assignee lists of the raw exports, names after a title such as ``Prof.'', and the name preceding a masked address), and usernames (submitter identifiers and the account list, plus the identifier after ``account'', ``username'' or ``login'', keeping English words, documentation terms, and project allocation identifiers). It masked 4{,}557 values, 3{,}773 of them usernames; afterwards no personal e-mail address or phone number remains in any split. Names absent from the exports and not marked by a title or an address can still slip through, which is one reason the data and the fine-tuned model are not released.

\textit{Split without leakage.} Exact-duplicate questions were dropped (48), leaving 12{,}644 pairs. Pairs were then assigned to splits \emph{by ticket}, with a fixed seed, so that the several pairs extracted from one ticket never straddle splits, and test and validation questions whose embedding (\texttt{all-MiniLM-L6-v2}) has cosine similarity $\geq$0.92 with any training question, or with another kept question of the same split, were removed, so that a user re-filing the same issue cannot place the test answer in the training data. The result is 11{,}244 training pairs, 600 validation pairs used only for validation loss and early stopping, and 400 test pairs used only for the evaluation in Section~\ref{sec:eval}. Rebuilding the split reproduces the files byte for byte; their hashes are checked before training.

\section{System}
\label{sec:system}

\begin{figure*}[t]
\centering
\includegraphics[width=0.92\textwidth]{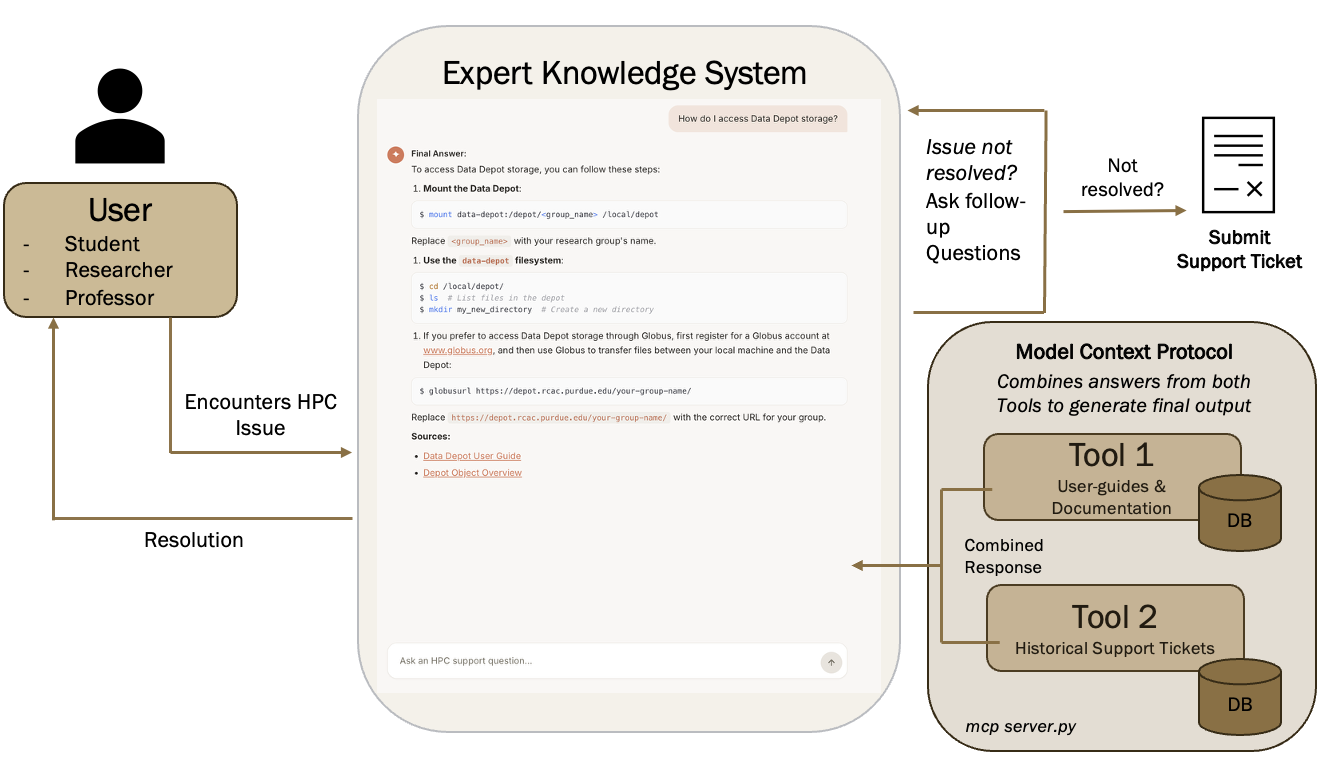}
\caption{Architecture. The orchestrator classifies each message; HPC-support queries invoke Tool~1 (RAG over user guides) and Tool~2 (model fine-tuned on resolved tickets) in parallel over MCP, and a synthesis step merges their answers with source links. If the answer does not resolve the issue, the assistant directs the user to the help desk, where they file a ticket as before.}
\label{fig:arch}
\end{figure*}

Figure~\ref{fig:arch} shows the design. We call it \emph{agentic} in the sense of \citet{sapkota2025agentic}: an LLM operates through an orchestration layer that decides how to handle each query and calls external tools rather than answering from parametric memory. The choice follows directly from the two-sources argument in the introduction. Keeping documentation and tickets in separate tools means each answer component stays grounded in its own material, and the orchestrator, not the training data, decides how to reconcile them.

A second constraint shaped every model choice. Support tickets contain names, paths, project identifiers, and occasionally credentials or grant details; even after de-identification, RCAC's data governance required that ticket text never be sent to an external API. Fine-tuning therefore had to fit on a single GPU of the center's own clusters, generation had to run on Purdue's institutionally hosted GenAI Studio (Llama~3.1), and embeddings had to be computed locally. Section~\ref{sec:discussion} notes where this limits comparison with frontier hosted models.

\subsection{Tool 1: RAG over the user guides}
A breadth-first crawler, seeded from the documentation home page and an explicit list of legacy pages on a second domain, collects content pages and discards images, assets, search and tag indices, and admin routes. 849 pages are fetched,and the main content is extracted as structured text: headings are kept, code blocks are fenced and never split, inline code stays inline, and line-number gutters and copy buttons are dropped. Each page is cut into chunks of whole paragraphs and code blocks from a single section, up to 900 characters, each prefixed with its section heading, giving 5{,}388 chunks. Chunks are embedded, with their page title, by a local Sentence-BERT model (\texttt{all-MiniLM-L6-v2}, 384-d) \citep{reimers2019sbert} into a FAISS flat index \citep{johnson2019faiss}, and also indexed with BM25.

For each question, candidates come from both the embedding search (top 40) and BM25 (top 40), so that an exact term such as \texttt{myquota} is found even when the embedding is dominated by a cluster name. A cross-encoder (\texttt{ms-marco-MiniLM-L-6-v2}) rescores every candidate. Pages documenting a cluster the question did not name are moved below all others, rather than boosting the named cluster, which had pushed relevant general pages out of the results; near-identical copies of one page across cluster guides are collapsed to one; and chunks below a relevance threshold are dropped. The threshold was set so that none of a sample of validation tickets lost all its chunks; when nothing remains the tool answers that the documentation does not cover the question, without calling the model (19 of the 400 test questions). The best six chunks go to Llama~3.1 on GenAI Studio, each labelled with its cluster and its relevance, with instructions to answer only from them, not to build an answer from low-relevance excerpts, to say so when they are not enough and point the user to human support, to ask a clarifying question when the request is ambiguous, and to cite the pages it used. A refresh endpoint re-crawls the documentation and swaps in a new index.

\subsection{Tool 2: what the tickets know}
We built two candidates and chose between them empirically.

\textbf{Fine-tuned model.} Llama-3-8B-Instruct \citep{llama3} was adapted with QLoRA \citep{dettmers2023qlora, hu2022lora} through Unsloth on a single NVIDIA A10 GPU of RCAC's Gilbreth cluster (Table~\ref{tab:ft}). Each example is a system/user/assistant exchange; the system message defines an RCAC support agent that gives user-facing guidance, never fabricates technical details, never claims to have performed back-end actions, asks for clarification when needed, and never asks for credentials. The serving configuration sends exactly the same system message. Loss is computed on assistant tokens only. A pilot sweep over $\{1,2,3\}\times10^{-4}$ on a sample of the training data found differences within noise, and we kept $2\times10^{-4}$, the QLoRA default. Validation loss on the 600-ticket split, measured every 200 steps, fell from 1.255 at step 200 to its minimum, 1.101 (perplexity 3.0), at step 2{,}800, the end of the second epoch. In the third epoch training loss kept falling, to about 0.70, while validation loss rose to 1.17--1.18, the model beginning to memorise its training resolutions; early stopping ended the run after three evaluations without improvement, at step 3{,}400. The step-2{,}800 checkpoint was merged, converted to GGUF, quantized to 4 bits (\texttt{q4\_K\_M}), and served locally with Ollama, so a tool call is a local inference request. The model receives only the question: it was trained on single question--answer pairs, so the orchestrator passes it the router's standalone rewrite of a follow-up rather than the conversation.

\begin{table}[t]
\centering\small
\caption{Fine-tuning configuration.}
\label{tab:ft}
\begin{tabular}{ll}
\toprule
Setting & Value \\
\midrule
Base model & Llama-3-8B-Instruct (4-bit) \\
Data & 11{,}244 train / 600 validation pairs \\
Adaptation & QLoRA; $r{=}16$, $\alpha{=}16$, dropout 0.05,\\
 & all projection layers \\
Max sequence length & 2048 \\
Batch & 2 $\times$ 4 grad.\ accumulation (effective 8) \\
Schedule & 1{,}406 steps per epoch; validation \\
 & every 200 steps; early stopping, \\
 & patience 3 evaluations \\
Training run & 2.4 epochs (stopped at step 3{,}400); \\
 & checkpoint used: step 2{,}800, \\
 & end of epoch 2 (val.\ loss 1.101) \\
Learning rate & $2\times10^{-4}$, linear, 5 warm-up steps \\
Optimiser & 8-bit AdamW, weight decay 0.001 \\
Loss / seed & Response tokens only / 3407 \\
Hardware & 1$\times$ NVIDIA A10 (24 GB), Gilbreth \\
Serving & GGUF \texttt{q4\_K\_M}, Ollama \\
\bottomrule
\end{tabular}
\end{table}

\textbf{RAG over tickets.} The same curated pairs become a retrieval corpus. Each training ticket is indexed by its extracted problem text with the same embedding model and FAISS setup as Tool~1; the top $k{=}5$ problem/resolution pairs go to Llama~3.1 with a prompt that permits only information from the retrieved resolutions, preserves technical tokens verbatim, and tells the user to file a ticket if nothing relevant was found. The index is built from the training split only, which the split construction keeps free of near-duplicates of the test questions.

\subsection{Orchestration over MCP}
Both tools are exposed through the Model Context Protocol, an open standard that declares each capability as a typed tool (name, description, JSON schema) discoverable by any MCP client. A FastMCP server publishes \texttt{query\_finetuned\_model} and \texttt{query\_rag\_docs} over stdio; the agent launches it as a subprocess and keeps a persistent client session. The decoupling is practical as much as architectural: the fine-tuned model and the RAG service were developed and redeployed independently throughout the project without touching the agent.

Each message is first handled by a lightweight routing prompt that sees the last few turns of the conversation. It decides whether the message is a technical question or conversation (greetings, thanks, questions about the assistant) and rewrites a follow-up such as ``what about on the GPU nodes?'' into a self-contained question before the tools see it; when in doubt, it treats the message as technical, and a message that names an RCAC cluster, service, or RCAC-specific command is never treated as conversation. Conversational messages are answered directly by Llama~3.1 with no tool calls. Technical messages trigger both tools concurrently (\texttt{asyncio.gather}), and a synthesis prompt merges the two outputs into one answer under an explicit rule: \emph{when the tools disagree on a concrete fact, a command, path, URL, prefer the documentation-grounded tool; otherwise, combine the useful parts of both answers.} Before the answer is shown, any link that did not come from one of the tools is removed, so every page the assistant cites is one it actually retrieved. A tool that fails reports a typed error rather than an answer; the synthesis then works from the other tool and the answer says which source was unavailable. All generation runs at temperature~0. When GenAI Studio does not respond, the same model tag is served from a local Ollama instance instead, so an outage degrades latency rather than availability.

\begin{figure}[t]
\centering
\includegraphics[width=\columnwidth]{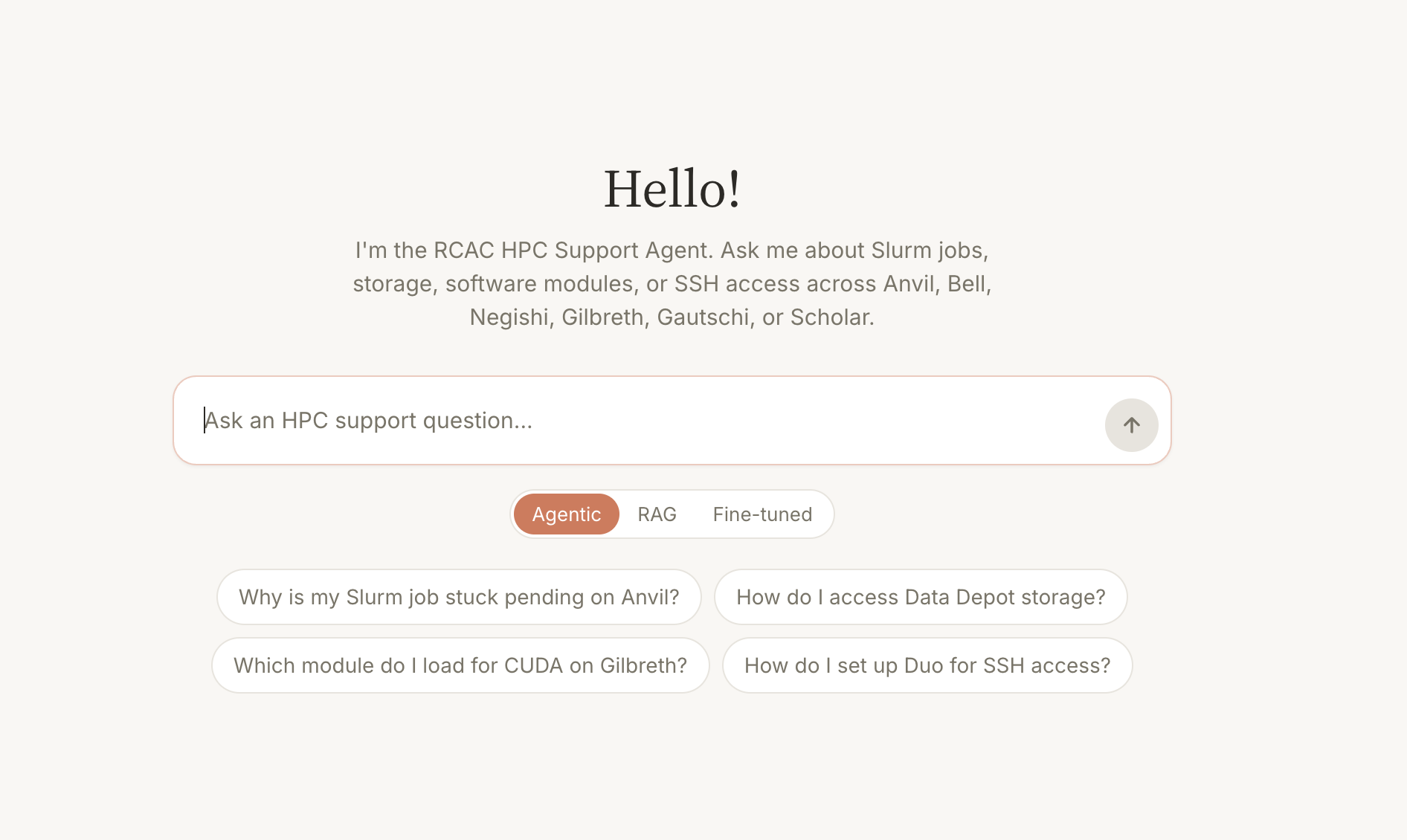}
\caption{The deployed interface. A mode selector (\emph{Agentic} / \emph{RAG} / \emph{Fine-tuned}) lets a user, or an evaluator, pose the same question to the full system or to either tool alone; the suggested prompts cover common kinds of support request.}
\label{fig:ui}
\end{figure}

The agent runs as a CLI and as a FastAPI service that streams classification, tool calls, and synthesis to a chat-style web UI over server-sent events (Figure~\ref{fig:ui}), so users see which sources were consulted and can follow a citation before running a command.

\section{Evaluation}
\label{sec:eval}

\subsection{Protocol}
\textbf{Systems and items.} Five systems answer each of the 400 test questions: the base model, Llama-3-8B-Instruct served exactly like the fine-tuned model (same 4-bit quantization, same system prompt, temperature~0, no adapter); the fine-tuned model (Tool~2); RAG over the documentation (Tool~1); RAG over the training tickets; and the end-to-end agentic system. Tool~1, Tool~2, and the agentic system are queried through the deployed chat API, the code path users reach, each question in a fresh session with no history, so that no answer sees another evaluation item. Every answer is saved together with the model digests, the split hashes, the index version, and the code version. All 2{,}000 answers were produced; no system failed on any item.

\textbf{Metrics.} Every answer is scored against the rewritten resolution of its test ticket. \textbf{BERTScore F1} \citep{zhang2020bertscore} (RoBERTa-large, baseline-rescaled) is the primary reference metric: it credits paraphrase but still compares content. \textbf{ROUGE-1/2/L F1} \citep{lin2004rouge} rewards exact tokens, which matters for commands and error strings, and penalises correct content absent from the reference. \textbf{Cosine similarity} between pooled \texttt{all-MiniLM-L6-v2} embeddings is reported as a secondary measure; it separates on-topic answers poorly. All three under-credit a correct answer that differs from the one recorded. An \textbf{LLM judge} \citep{zheng2023judging}, given question, reference, and candidate, scores 1--5 on \emph{correctness} (consistent with the reference; invented commands, paths, modules, URLs, or claims that staff performed an action are penalised), \emph{completeness} (necessary steps and caveats present), and \emph{helpfulness} (could the user resolve the issue, regardless of wording), and is told that length is not a merit.

\textbf{Judges.} Every answer is judged by two models from families that generated none of the answers: Qwen2.5-72B and Gemma-3-27B, both on GenAI Studio, at temperature~0. The judge never sees which system wrote an answer, items are judged in shuffled order, and every score is saved; 1{,}999 and 1{,}998 of the 2{,}000 answers were scored. We report the mean of the two judges and each judge separately. All means come with 95\% bootstrap intervals over items, and comparisons with Tool~2 use paired bootstrap intervals on the same items; an interval that contains zero is reported as no difference.

\subsection{Results}
Table~\ref{tab:tool2} compares the Tool~2 candidates and Table~\ref{tab:overall} compares Tool~1, Tool~2, and the agentic system, on the same 400 items, together with an oracle that picks the better of the two tool answers for each question. Table~\ref{tab:judges} reports judged correctness for each judge separately.

\begin{table}[t]
\centering\small
\caption{Tool~2 candidates on the same 400 held-out test tickets (higher is better; judge scores 1--5, mean of two judges). Bold: best in row. For every reference and judge metric, the difference between the fine-tuned model and each other candidate has a 95\% paired-bootstrap interval that excludes zero.}
\label{tab:tool2}
\begin{tabular}{lccc}
\toprule
Metric & Base & Fine-tuned & Ticket-RAG \\
\midrule
BERTScore F1 & 0.061 & \textbf{0.269} & 0.118 \\
ROUGE-1 F1 & 0.281 & \textbf{0.435} & 0.341 \\
ROUGE-2 F1 & 0.082 & \textbf{0.165} & 0.115 \\
ROUGE-L F1 & 0.156 & \textbf{0.271} & 0.191 \\
Cosine similarity & 0.704 & \textbf{0.736} & 0.698 \\
\midrule
Judge: correctness & 2.87 & \textbf{3.20} & 2.86 \\
Judge: completeness & 3.07 & 2.85 & \textbf{3.11} \\
Judge: helpfulness & \textbf{3.66} & 3.57 & 3.42 \\
\midrule
Answer length (words) & 255 & 76 & 184 \\
\bottomrule
\end{tabular}
\end{table}

\begin{table}[t]
\centering\small
\caption{Answer quality for Tool~1 (RAG on docs), Tool~2 (fine-tuned), and the end-to-end agentic system on the same 400 held-out test tickets (judge scores 1--5, mean of two judges). Bold: best in row among the three systems. Differences from Tool~2 have 95\% paired-bootstrap intervals that exclude zero, except Tool~1's completeness ($-0.05$, $[-0.13, 0.04]$). \emph{Best of two} is an oracle, not a system: for each question it takes whichever tool answer the judges rated more correct (Tool~1 on 104 questions, Tool~2 otherwise, ties to Tool~2) and scores that answer; because it selects on the same judge scores, its judge rows are an optimistic upper bound for choosing between the tools.}
\label{tab:overall}
\setlength{\tabcolsep}{4pt}
\begin{tabular}{lcccc}
\toprule
Metric & Tool~1 & Tool~2 & Agentic & \textit{Best of two} \\
\midrule
BERTScore F1 & 0.021 & \textbf{0.269} & 0.041 & \textit{0.225} \\
ROUGE-1 F1 & 0.302 & \textbf{0.435} & 0.313 & \textit{0.416} \\
ROUGE-2 F1 & 0.079 & \textbf{0.165} & 0.082 & \textit{0.159} \\
ROUGE-L F1 & 0.165 & \textbf{0.271} & 0.175 & \textit{0.257} \\
Cosine similarity & 0.659 & \textbf{0.736} & 0.649 & \textit{0.734} \\
\midrule
Judge: correctness & 2.78 & \textbf{3.20} & 2.66 & \textit{3.41} \\
Judge: completeness & 2.81 & \textbf{2.85} & 2.63 & \textit{2.99} \\
Judge: helpfulness & 3.24 & \textbf{3.57} & 3.20 & \textit{3.68} \\
\midrule
Answer length (words) & 154 & 76 & 112 & \textit{98} \\
\bottomrule
\end{tabular}
\end{table}

\begin{table}[t]
\centering\small
\caption{Judged correctness (1--5) by each judge, and the paired difference Tool~2 (fine-tuned) minus each system with its 95\% bootstrap interval; $n{=}400$ per system. Agreement between the judges over all 2{,}000 answers (Spearman $\rho$): correctness 0.63, completeness 0.62, helpfulness 0.54.}
\label{tab:judges}
\resizebox{\columnwidth}{!}{%
\begin{tabular}{lcccc}
\toprule
& \multicolumn{2}{c}{Qwen2.5-72B} & \multicolumn{2}{c}{Gemma-3-27B} \\
\cmidrule(lr){2-3}\cmidrule(lr){4-5}
System & Mean & Tool~2 $-$ system & Mean & Tool~2 $-$ system \\
\midrule
Base & 3.34 & $-0.13$ $[-0.22, -0.05]$ & 2.40 & $+0.79$ $[+0.66, +0.91]$ \\
Tool~2 (fine-tuned) & 3.21 & --- & 3.19 & --- \\
Tool~1 (docs RAG) & 3.18 & $+0.03$ $[-0.06, +0.12]$ & 2.38 & $+0.81$ $[+0.69, +0.93]$ \\
Ticket-RAG & 3.15 & $+0.05$ $[-0.04, +0.14]$ & 2.58 & $+0.62$ $[+0.49, +0.74]$ \\
Agentic & 3.00 & $+0.21$ $[+0.11, +0.31]$ & 2.32 & $+0.87$ $[+0.76, +0.99]$ \\
\bottomrule
\end{tabular}}
\end{table}

\textbf{Fine-tuning wins on agreement with staff resolutions.} Over the base model served the same way, the adapter raises BERTScore F1 from 0.06 to 0.27, ROUGE-1 from 0.28 to 0.44, ROUGE-L from 0.16 to 0.27 and cosine from 0.70 to 0.74, all with intervals far from zero, and it does so with answers less than a third as long (76 vs.\ 255 words). The model absorbed RCAC's vocabulary, response format, and recurring fixes. Part of this lead is built in: the model was trained to reproduce the register of exactly the kind of text it is scored against, and the reference answers are themselves LLM rewrites of staff replies. The judges, which do not reward register, give a more mixed picture (below).

\textbf{Retrieval over the same tickets is the weaker way to use them, but writes the most complete answers.} Ticket-RAG falls well short of the fine-tuned model on every reference metric (BERTScore 0.12 vs.\ 0.27) and on judged correctness ($-0.33$) and helpfulness ($-0.15$), but it is judged the most complete system of all (3.11 vs.\ 2.85), and against the base model it ties on correctness and completeness. It paraphrases the resolutions of similar but different tickets into long, step-by-step answers, which carry more steps but more that do not apply. Ticket-RAG keeps one advantage the snapshot cannot show: it is updated by re-indexing rather than retraining.

\textbf{Documentation answers a different question than the reference does.} Tool~1 has the lowest BERTScore (0.02) and cosine of the tool candidates, and it is judged below Tool~2 on correctness ($-0.42$) and helpfulness ($-0.33$), with a tie on completeness. This is partly the protocol: the references record what support staff did for one user, often an account or allocation change that no user guide describes, while Tool~1 answers from the current guides, and declined to answer 19 questions the guides do not cover. A correct answer from the documentation can still score low here.

\textbf{The end-to-end system does not improve on its tools.} The agentic system is below Tool~2 on every metric and under both judges (correctness $-0.54$ on the mean, $-0.21$ by Qwen, $-0.87$ by Gemma). Against Tool~1 it is marginally ahead on the reference metrics (BERTScore $+0.02$, $[+0.01, +0.03]$), which its shorter answers (112 vs.\ 154 words) and some of Tool~2's ticket vocabulary account for, but behind on judged correctness ($-0.12$, $[-0.20, -0.04]$) and completeness ($-0.18$, $[-0.26, -0.10]$), with helpfulness tied. In substance, then, it improves on neither tool. The router accounts for little of this: 13 of 400 questions (3\%), all genuine support requests about accounts, storage folders, training materials or services, were answered as conversation without tools, and excluding them leaves the agentic correctness unchanged (2.67 vs.\ 2.66).

\textbf{The tools are complementary; the merge loses it.} The architecture's premise holds: each tool clearly wins on part of the test set, by at least one point of judged correctness, 144 questions for Tool~2 and 44 for Tool~1; on the remaining 212 they are within a point of each other. The synthesis does not exploit this. Its correctness stays near 2.65 whatever its inputs: 2.68 where Tool~2 is clearly better (Tool~2: 3.95), 2.64 where Tool~1 is clearly better (Tool~1: 3.52), and 2.64 where the two are about equal (2.83 and 2.88), so merging adds errors even when the tools agree, and on 30\% of questions the merged answer is at least half a point below the \emph{worse} of its two inputs. Part of the reason is the precedence rule: on the 387 items routed to the tools, the merged answer overlaps more with Tool~1's (ROUGE-L 0.27) than with Tool~2's (0.19) and is closer to Tool~1's in 78\% of items, so ``prefer the documentation on concrete facts'' pulls the answer toward the tool that is right less often on ticket-style questions. Choosing the better tool answer per question (Table~\ref{tab:overall}, \emph{Best of two}) would raise judged correctness to 3.41 ($+0.21$ over Tool~2, $[+0.17, +0.26]$; $+0.28$ by Qwen, $+0.15$ by Gemma), completeness to 2.99 and helpfulness to 3.68, at a small cost in reference overlap (BERTScore $-0.04$). This oracle selects on the judge scores it is evaluated on, so it is an upper bound, not an achievable result; it shows that the step to improve is the merge, not the tools.

\begin{table}[t]
\centering\small
\caption{Choosing one tool answer per question instead of merging, on the 400 test tickets (judge scores 1--5; $\Delta$: paired difference from Tool~2 alone, 95\% bootstrap CI). Both selectors were developed on the 600-ticket validation split only and see the question and the two answers, never a reference. \emph{Picks docs}: share of questions answered with Tool~1's answer. The oracle row repeats Table~\ref{tab:overall}'s upper bound.}
\label{tab:select}
\resizebox{\columnwidth}{!}{%
\begin{tabular}{lccccccc}
\toprule
& Picks & \multicolumn{2}{c}{Correctness} & Compl. & Help. & \multicolumn{2}{c}{Correctness by} \\
\cmidrule(lr){3-4}\cmidrule(lr){7-8}
Strategy & docs & Mean & $\Delta$ vs.\ Tool~2 & & & Qwen & Gemma \\
\midrule
Merge (agentic) & --- & 2.66 & $-0.54$ $[-0.64, -0.45]$ & 2.63 & 3.20 & 3.00 & 2.32 \\
Tool~2 alone & 0\% & 3.20 & --- & 2.85 & 3.57 & 3.21 & 3.19 \\
Learned selector & 6\% & 3.19 & $-0.01$ $[-0.02, +0.01]$ & 2.87 & 3.57 & 3.22 & 3.16 \\
LLM verifier & 40\% & 3.08 & $-0.11$ $[-0.17, -0.05]$ & 2.98 & 3.54 & 3.26 & 2.91 \\
\midrule
\textit{Oracle} & \textit{26\%} & \textit{3.41} & \textit{$+0.21$ $[+0.17, +0.26]$} & \textit{2.99} & \textit{3.68} & \textit{3.49} & \textit{3.33} \\
\bottomrule
\end{tabular}}
\end{table}

\textbf{Selecting instead of merging recovers Tool~2's quality, but no realisable selector beats Tool~2.} We tested the obvious repair, returning one tool's answer unchanged, with two selectors that see only what the live agent has, the question and the two answers (Table~\ref{tab:select}). The \emph{learned selector} is a logistic regression over the question's words and signals from the answers (the documentation tool abstaining or saying the guides do not cover the question, the fine-tuned answer routing the user to staff, answer lengths), with its regularisation and decision threshold chosen by cross-validation on validation. The \emph{LLM verifier} asks the agent's own model, Llama~3.1, which answer better resolves the question, once in each order, and keeps Tool~2's answer when the orders disagree. Either is far better than the merge ($+0.52$, $[+0.43, +0.62]$, and $+0.40$ judged correctness), but neither beats Tool~2 alone. The learned selector found little usable signal and picks the documentation answer for only 6\% of questions; its strongest cues are plausible (modules, \texttt{sbatch}, conda or nodes lean to the documentation; home directories, scratch, files or access to the tickets) but weak. The verifier picks the documentation answer for 40\% of questions, more often than the oracle's 26\%; on validation it chose the judges' clear winner in only 63\% of 285 clear cases, below the 81\% of always choosing Tool~2, and its two orders agreed on only 51\% of questions, a strong position bias. It gains completeness ($+0.14$) by picking longer answers, at a cost in correctness ($-0.11$) that the judges again see differently ($+0.02$ by Qwen, $-0.30$ by Gemma). Knowing in advance which source will be right is itself a hard problem.

\textbf{The judges disagree about correctness.} Averaged, the judges rate Tool~2 the most correct system. Separately they tell different stories (Table~\ref{tab:judges}). Gemma-3-27B rates Tool~2 well above every other system, by 0.62--0.87 points. Qwen2.5-72B rates the base model slightly \emph{above} Tool~2 ($-0.13$) and finds no difference between Tool~2 and either RAG system; it is also more lenient overall (38\% of answers scored 4--5 for correctness, against 16\% for Gemma) and spreads its scores less. The two agree only moderately (Spearman $\rho$ 0.63 for correctness, 0.54 for helpfulness). On completeness and helpfulness Qwen again prefers the base model ($-0.38$ and $-0.35$ for Tool~2), while Gemma finds no completeness difference and rates Tool~2 more helpful ($+0.15$). Both judges agree that the agentic system is the least correct. The claim that fine-tuning improves correctness, as opposed to agreement with the reference, therefore depends on the judge.

\textbf{Completeness rewards length.} Across all 2{,}000 answers, answer length correlates with judged completeness (Spearman $\rho{=}0.37$) and helpfulness (0.15) but not with correctness ($-0.04$), and negatively with BERTScore ($-0.40$), even though the judge prompt says length is not a merit. The base model (255 words) and ticket-RAG (184 words) lead on completeness largely by writing more. Completeness and correctness should be read together; a system that wins only on completeness has mostly written more.

\section{Discussion and Limitations}
\label{sec:discussion}

\textbf{Treat tickets as data, not as answers.} The largest quality risk we encountered was not hallucination from the model but faithful reproduction of the wrong thing: resolutions in which a human did something the assistant cannot do. Rewriting those into ``here is how to request this'' guidance, and deleting status updates that were true for an afternoon, removed a whole class of confidently false answers, and a first, narrower filter still let 3\% through. The same holds for privacy: pseudonymisation that looked complete still left thousands of usernames that a fine-tuned model can learn and repeat. Anyone fine-tuning on help-desk history should expect to iterate on both.

\textbf{Select, do not merge.} Fine-tuning beat retrieval over the same tickets on every reference metric and, by one of two judges, on correctness, with answers a third the length. The orchestration failed in a specific place: the sources are complementary, but a rule that makes the documentation win on concrete facts made the synthesis track the documentation tool, and rewriting two answers into one lost content even when both were good. Selecting one answer recovers the fine-tuned tool's quality but not the oracle's headroom, so if one must choose today, returning the fine-tuned answer, with the documentation answer offered alongside as a cited second opinion, is safer than any merge we tried. Better selection signals, such as the retrieval relevance of the documentation evidence, a check of each answer's commands against the cluster the question names, or user feedback, are the next step, as is a hybrid Tool~2 that retrieves recent tickets to stay current.

\textbf{Report every judge and read dimensions together.} Two capable judges from different model families agreed on the weakest system but disagreed on whether fine-tuning improved correctness, and both rewarded length on completeness. A single judge, or an average without its parts, would have produced a confident and judge-specific conclusion, so we report each judge and size our claims to what holds under both.

\textbf{Limitations.} All scores come from reference metrics and LLM judges, which disagree on the central correctness comparison and have not been validated against human ratings for this task. An expert evaluation is under way and will be added in a revision: RCAC support staff rate a blinded sample of test tickets, the staff resolution and four systems' answers per question, for correctness, completeness, and whether they would send the answer to the user. The references are LLM-rewritten staff resolutions, the kind of text the fine-tuned model was trained on, which favours it on reference metrics; many concern account, allocation, or storage changes that only staff can make, which the documentation tool cannot answer from the guides. The selectors were tuned on validation labels from the same two judges, so they learn the judges' preferences; the oracle uses test labels and is an upper bound, not a system. Because GenAI Studio rate-limited heavily, a local copy of the same model tag (\texttt{llama3.1:latest} on Ollama) served part of the generation calls (643 of the documentation service's, 149 of the agent's, 51 of the 400 ticket-RAG answers, and 1{,}700 of the verifier's), though no judge scores. GenAI Studio exposes model tags, not digests, so the hosted models can change underneath; the fine-tuned and base models are pinned by digest. No frontier model was compared, because ticket text could not leave the institution; the public documentation would allow such a comparison for Tool~1. 

\section{Related Work}
\label{sec:related}

\textbf{LLM assistants for HPC user support.} This is now an active line at HPC-operations venues. AskHPC \citep{bondapalli2025askhpc} is a RAG chatbot over user guides, scheduler manuals, and programming documentation at a leadership facility, with modality-aware parsing that keeps code blocks intact and a user study showing gains over vanilla RAG; it does not use ticket history. HPC-LLM \citep{shahin2026hpcllm} pairs QLoRA domain adaptation with retrieval, as we do, but trains on public documentation and synthetic instruction pairs rather than real resolutions.

Closest to our evaluation, \citet{zheng2026comparative} compare fine-tuning on historical tickets, RAG over documents, and in-context examples in eight combinations on roughly 60{,}000 TACC tickets, scoring accuracy and coherence. Relative to that work, we keep documentation and tickets in separate orchestrated tools rather than one prompt or one adapter, show that rewriting and filtering staff-performed resolutions is a prerequisite for a deployable assistant, and find that a simple precedence rule for merging the two sources does not combine their strengths.

\textbf{Fine-tuning versus retrieval.} \citet{ovadia2023finetuning} find that RAG consistently outperforms unsupervised fine-tuning for injecting new factual knowledge, while \citet{balaguer2024ragvsft} find in an agricultural case study that the two are complementary and that fine-tuning chiefly buys style and domain fluency. Our results sit between them: fine-tuning beat retrieval over the same tickets on agreement with staff resolutions, consistent with capturing \emph{how} RCAC answers, and fine-tuned and documentation answers were complementary in \citeauthor{balaguer2024ragvsft}'s sense, but merging them did not realise that complementarity.

\textbf{Agentic and tool-using LLMs.} ReAct \citep{yao2023react} established the loop of reasoning interleaved with tool calls; agentic RAG surveys \citep{singh2025agenticrag} catalogue orchestrator-plus-tool designs in which retrieval is one capability among several; \citet{sapkota2025agentic} distinguish single tool-augmented agents from multi-component agentic systems. The Model Context Protocol \citep{mcp2024} standardises the tool interface we use. Our orchestrator is deliberately simple, one classification, two parallel tool calls, one synthesis; our results suggest the synthesis is where such a design needs the most care.

\textbf{Ticket automation.} Surveys find a fragmented literature of sub-task automation, classification, routing, escalation, mostly with classical ML \citep{fuchs2022, zangari2023}. LLMs now handle escalation \citep{liu2025}, categorisation \citep{paket2024}, and RAG-based customer service \citep{xi2024, xu2024}, building on \citet{lewis2020}. All of these act after a ticket exists; we act before.

\textbf{Documentation usability.} Inadequate documentation is a primary obstacle when learning new systems \citep{robillard2011}, and HPC users lean on facilitators and peers over written guides \citep{samuel2021, suleman2025, raj2020}; these studies motivate contextual delivery but do not evaluate guides and ticket history as one system.

\textbf{LLM-as-judge.} \citet{zheng2023judging} validate LLM judges against human preference but document position and verbosity biases; \citet{panickssery2024selfpref} show judges favour text from their own model family, which is why our judges come from families that generated none of the answers. We observe both a verbosity effect on completeness and a substantive disagreement between two such judges.

\section{Conclusion}
\label{sec:conclusion}

The knowledge needed to answer most HPC support tickets already exists, split between manuals that are current but generic and a ticket history that is specific but inaccessible. Within the constraints of on-premises models, fine-tuning a small model on carefully rewritten and filtered resolutions is an effective way to use that history: on 400 held-out tickets it agreed with RCAC's resolutions far better than the base model or retrieval over the same tickets, and one of two judges rated it clearly the most correct. The two sources are complementary, but merging them under a documentation-first rule improved on neither, and selecting one answer recovers the fine-tuned tool's quality without exceeding it: deciding in advance which source will be right is the open problem. The steps most responsible for a trustworthy assistant were data work: rewriting staff actions into self-service guidance, filtering what rewriting missed, and masking what pseudonymisation missed. Validating the judges against expert ratings, to be reported in a revision, and better signals for choosing the source are the next steps.

\paragraph{Acknowledgements.}
% TODO: edit as appropriate.
We thank the Rosen Center for Advanced Computing at Purdue University for providing data, technical guidance, and domain expertise.

\small
\bibliographystyle{plainnat}
\bibliography{refs}

\end{document}